\documentclass[letterpaper]{JHEP3}
\input psfig
\usepackage{citesort}
\newcount\figno
\def\fig#1#2#3{
\par\begingroup\parindent=0pt\leftskip=1cm\rightskip=1cm\parindent=0pt
\global\advance\figno by 1
\midinsert
\epsfxsize=#3
\centerline{\epsfbox{#2}}
\vskip 12pt
{\bf Fig. \the\figno:} #1\par
\endinsert\endgroup\par
}
\def\figlabel#1{\xdef#1{\the\figno}}
\def\encadremath#1{\vbox{\hrule\hbox{\vrule\kern8pt\vbox{\kern8pt
\hbox{$\displaystyle #1$}\kern8pt}
\kern8pt\vrule}\hrule}}
\def\underarrow#1{\vbox{\ialign{##\crcr$\hfil\displaystyle
 {#1}\hfil$\crcr\noalign{\kern1pt\nointerlineskip}$\longrightarrow$\crcr}}}
\def\Tr{{\rm Tr}}
\def\be{\begin{equation}}
\def\ee{\end{equation}}
\def\tilde{\widetilde}
\def\bar{\overline}

\def\T{{\bf T}}

\def\R{{\bf R}}

\font\zfont = cmss10 
\font\litfont = cmr6

\def\bigone{\hbox{1\kern -.23em {\rm l}}}
\def\ZZ{\hbox{\zfont Z\kern-.4emZ}}
\def\half{{\litfont {1 \over 2}}}

\title{BPS Bound States of $D0\,$-$D6$ and $D0\,$-$D8$ Systems in a $B$-Field}
\author{Edward Witten\thanks{Currently, at School of Natural Sciences, Institute for Advanced Study, Princeton, NJ 08540}\\ Dept. of Physics, Cal Tech, Pasadena, CA\\ CIT-USC Center for Theoretical Physics, USC, Los Angeles, CA\\ E-mail: \email{witten@ias.edu}}
\abstract{The $D0\,$-$D6$ system, which is not supersymmetric in the absence of
a Neveu-Schwarz $B$-field, becomes supersymmetric if a suitable 
constant $B$-field is turned on.  On one side of the supersymmetric
locus, this system has a BPS bound state, and on the other side it does
not.   After compactification on $\T^6$, this gives a simple example in which
the number of 1/8 BPS states jumps as the moduli of the compactification
are changed.  The $D0\,$-$D8$ system in a $B$-field has two different 
supersymmetric loci, only one of which is continuously connected to
the familiar supersymmetric $D0\,$-$D8$ system without a $B$-field.
In a certain range, the $D0\,$-$D8$ system also has a BPS bound state.
In the limit in which the $B$-field goes to infinity, supersymmetric
$D0\,$-$D6$ and $D0\,$-$D8$ systems and their bound states can be studied using noncommutative Yang-Mills theory.}
\preprint{hep-th/0012054}
\begin{document}
\section{Introduction}

$D0\,$-$Dp$ brane systems in Type IIA superstrings
(and more general $Dq\,$-$Dp$ systems) have been much studied.
Some aspects have been reviewed in  \cite{polch}.  In the absence of a $B$-field, the $D0\,$-$D0$, $D0\,$-$D4$,
and $D0\,$-$D8$ systems are supersymmetric, while  the others are not.
In the presence of a constant $B$-field, the condition for supersymmetry
is modified.  For example, the $D0\,$-$D4$ system in a constant $B$-field
is supersymmetric precisely if the $B$-field is anti-self-dual
\cite{sw}.
More general conditions for supersymmetry in the presence of a $B$-field
have been discussed in \cite{chen,inyong}.

The $D0\,$-$D2$ system
remains  non-supersymmetric in the presence
of a $B$-field.   However, as we will review in section 2, the $D0\,$-$D6$
system becomes supersymmetric when a suitable $B$-field is turned on.
In the space of constant $B$-fields, there is a codimension one locus
on which there is unbroken supersymmetry.  In section 3, we show that
on one side of the supersymmetric locus, the $D0\,$-$D6$ system has
a supersymmetric or BPS bound state, invariant under 1/8 of the
total supersymmetry; on the other side it does not.
This result still holds if the $D0\,$-$D6$ system is compactified on $\T^6$;
it demonstrates that in Type IIA superstring theory compactified on $\T^6$,
the number of 1/8 BPS states can jump as the moduli of the vacuum are changed.
This jumping is analogous to the jumping of 1/4 BPS states
in ${\cal N}=4$ super Yang-Mills theory in four dimensions
\cite{bergman}.

The $D0\,$-$D8$ system, on the other hand,
is supersymmetric in the absence of a
$B$-field.  Turning on the $B$-field can preserve supersymmetry,
and in addition there is a second supersymmetric
locus of the $D0\,$-$D8$ system, not continuously connected to the
theory with vanishing $B$-field.  In a certain range, there is
a supersymmetric $D0\,$-$D8$ bound state.

It is possible to take the $B$-field to infinity and obtain
supersymmetric $D0\,$-$D6$ and $D0\,$-$D8$ systems that can be described
in noncommutative Yang-Mills theory, as in \cite{cds,sw}.
The relevant solutions of noncommutative Yang-Mills theory
can be obtained by straightforwardly 
imitating arguments that have been given in the
recent literature for the $D1\,$-$D3$, $D0\,$-$D2$, and $D0\,$-$D4$ cases
\cite{gn,polyc,mand,bak,stromet,harveyet,gntwo}.  Presumably, in a suitable range
of $B$-fields where a BPS bound state exists, the noncommutative
Yang-Mills theory could be used to demonstrate the decay of the $D0\,$-$D6$
and $D0\,$-$D8$ systems to a stable bound state, similarly to 
discussions of $D0\,$-$D2$ in \cite{stromet,gntwo}.

The analysis in section 4 is largely anticipated in \cite{inyong},
as I learned after submitting the original version of this paper
to hep-th.  In addition, the problem is $T$-dual to a problem treated
in \cite{kachru} involving $D3$-branes intersecting at angles
\cite{douglas}
\section{Supersymmetric $D0\,$-$D6$ and $D0\,$-$D8$ Systems}

In Type IIA superstring theory, the supercharges that originate
from left- and right-movers on the string worldsheet transform
as spinors $Q_\alpha$ and $\tilde Q^\beta$, respectively, of opposite
chirality.  For a linear combination $\sum_\alpha \epsilon^\alpha Q_\alpha
+\sum_\beta \tilde \epsilon_\beta \tilde Q^\beta$ to be unbroken
in the presence of a $D$-brane, a certain condition must be obeyed,
depending on the brane.  

For example, in the field of a $D0$-brane at rest, with $x^0$ being
the time direction and $x^1,\dots,x^9$ the space directions, the condition is
\be
\tilde\epsilon_\beta =\Gamma^0_{\beta\alpha}\epsilon^\alpha,
\label{nuggo}
\ee
where $\Gamma^i$ are Dirac gamma matrices.
This equation has a simple (and well known; see for instance
\cite{polch}) interpretation.  Consider an open string that
ends on the $D0$-brane.  Because the worldsheet field $X^0$ corresponding
to $x^0$ obeys Dirichlet boundary conditions
and the other fields
$X^i$, $i>0$, obey Neumann boundary conditions, the worldsheet
modes when reflected from the end of the string are multiplied by a matrix
that (in the vector representation of the ten-dimensional Lorentz
group) is ${\rm diag}(1,-1,-1,\dots,-1)$.  In the spinor representation,
this matrix is $\Gamma^0$.

Now consider a $D6$-brane whose world-volume fills the directions
$x^0,x^1,\dots,x^6$, but with vanishing $B$-field.
 The condition for unbroken supersymmetry in the field of such a brane is
\be
\tilde\epsilon_\beta=(\Gamma^0\Gamma^1\cdots\Gamma^6)_{\beta\alpha}
\epsilon^\alpha.
\label{tuggo}
\ee
This reflects the fact that for a string ending on the $D6$-brane,
$X^0,\dots,X^6$ obey Dirichlet boundary conditions and the others
obey Neumann boundary conditions, so the reflection matrix is
${\rm diag}(-1,-1,\dots,-1,1,1,1)$, with $-1$'s for $x^0,\cdots,x^6$.

In the presence of both a $D0$-brane and a $D6$-brane, the unbroken
supersymmetries must obey both conditions.
Combining them, we get
\be
(\Gamma^1\Gamma^2\cdots\Gamma^6)\epsilon=\epsilon.
\label{kugo}
\ee
This equation has no nonvanishing solutions, since $N=\Gamma^1\Gamma^2\cdots
\Gamma^6$ obeys $N^2=-1$ and has all eigenvalues $\pm i$.
This shows that, without the $B$-field, the $D0\,$-$D6$ system is not
supersymmetric.  

As the above explanation makes clear, 
the matrix $N$ that appears here has a simple interpretation: it 
represents the action on spinors of the 
Lorentz transformation that acts as $-1$ on $x^1,\dots ,x^6$ and $+1$
on the other coordinates.  We can think of this group element as an element
 of the center of the
rotation group $SO(6)$ that acts on $x^1,\dots,x^6$.

Now, let us turn on a constant $B$-field in the directions $x^1,\dots,x^6$.
Consider a string that ends on the $D6$-brane in the presence of the
$B$-field.
In the RNS description, the
 worldsheet bosons and fermions $x^i,\psi^i$, $i=1,\dots,6$, when they are reflected from the string
end,
undergo a rotation by an element $M$ of $SO(6)$ that depends on $B$.
We can pick a coordinate system in which $B$ is the sum of three $2\times 2$
blocks, with the $i^{th}$ such block taking the form
\be
{1\over 2\pi\alpha'}\left(\matrix{ 0 & -b_i \cr b_i & 0 \cr}\right).
\label{omo}
\ee
Then \cite{tf,napetal} $M$ is likewise the sum of three $2\times 2$ blocks,
with the $i^{th}$ block  being
\be
\left(\matrix{\cos\,2\pi v_i&\sin\,2\pi v_i\cr -\sin\,2\pi v_i&\cos\,2\pi v_i}\right),
\label{jomo}
\ee
with
\be
e^{2\pi iv_i}={1+ib_i\over 1-ib_i},~~-\half <v_i<\half.
\label{hoko}
\ee
($v_i$ has here been shifted by $1/2$ relative to \cite{sw}.)

When we allow for this extra rotation matrix, 
the worldsheet modes of a string ending on a $D6$-brane are rotated
not just by the Lorentz transformation that in the spinor representation
is $\Gamma^0\Gamma^1\cdots \Gamma^6$; there is an extra factor $\rho(M)$,
where $\rho(M)$ is simply $M$ written in the spinor representation.
(\ref{tuggo}) becomes
\be
\tilde\epsilon_\beta=(\Gamma^0\Gamma^1\cdots\Gamma^6\rho(M))
_{\beta\alpha}
\epsilon^\alpha,
\label{otuggo}
\ee
and (\ref{kugo}) becomes
\be
(\Gamma^1\Gamma^2\cdots\Gamma^6\rho(M))\epsilon=\epsilon,
\label{okugo}
\ee
or more simply
\be
\rho(-M)\epsilon=\epsilon,
\label{nokuho}
\ee
\def\C{{\bf C}} since $\Gamma^1\Gamma^2\cdots\Gamma^6=\rho(-1)$ represents
$-1$ in the spinor representation.

Thus, the condition for unbroken supersymmetry is that the $SO(6)$
element $-M$ must, in the spinor representation, have $+1$ as one of
its eigenvalues.  The condition for this is familiar: $-M$ must be an element
of some $SU(3)$ subgroup of $SO(6)$.  
If we complexify the space generated by $x^1,\dots, x^6$ to make
a copy of $\C^6$, then the $SU(3)$ will leave fixed a three-dimensional
subspace $\C^3\subset \C^6$.  This $\C^3$ must be generated by eigenspaces
of the matrix $M$.
If the three blocks in (\ref{omo})
are assumed to act on the 1-2, 3-4, and 5-6 planes, 
then the eigenspaces of $M$
are generated by  $x^1\pm ix^2$, $x^3\pm ix^4$, and $x^5\pm ix^6$. The relevant
$\C^3$ is generated by $x^1\pm i x^2$, $x^3\pm i x^4$, and $x^5\pm i x^6$ with
some specific choices of the signs.  This $\C^3$ is invariant under a $U(3)$ that contains $-M$; the condition that $-M$ is actually in $SU(3)$ is that,
if it is understood as a $3\times 3$ diagonal matrix with eigenvalues
$-e^{\pm 2\pi i v_k}$, $k=1,2, 3$ (with some choices of sign), its determinant must be 1.  The
condition in other words is that
\be
\pm v_1\pm v_2\pm v_3= \pm 1/2,
\label{ono}
\ee
with some choices of the signs.
In fact, by possibly reversing orientation in the 1-2, 3-4, and 5-6 planes,
we can always arrange so that the signs are all positive, and the condition
for unbroken supersymmetry is
\be
v_1+v_2+v_3=1/2.
\label{ptono}
\ee
When this condition is obeyed and the $v_i$ are otherwise generic,
(\ref{nokuho}) is obeyed for one-fourth of the components of $\epsilon$.
Allowing also for (\ref{otuggo}), one-eighth of the 32 supersymmetries are unbroken.
Thus, there are four unbroken supersymmetries in all.

\bigskip\noindent
{\it Comparison To $D0\,$-$D4$}

Let us now compare this to the familiar result for $D0\,$-$D4$.
  For $D0\,$-$D4$, the matrix $-M$
has eigenvalues $-e^{\pm 2\pi i v_k}$, $k=1,2$.  The condition
for its determinant to be 1 as a $U(2)$ matrix is
\be
v_1\pm v_2=0.
\label{tono}
\ee
With the basis $x^1,\dots,x^4$ oriented in the fashion assumed
in \cite{sw}, the sign is $+$.  (\ref{tono}) together with (\ref{jomo})
means that $B_1=-B_2$ so that
the $B$-field is anti-self-dual.

(\ref{tono}) implies that the eigenvalues of $-M$ are not distinct,
so when (\ref{tono}) holds, $-M$  commutes with a subgroup of $SO(4)$
of larger than the generic dimension and is on a conjugacy class in $SO(4)$
smaller than the generic dimension.  Indeed, 
when (\ref{tono}) is obeyed, $-M$ takes values in one $SU(2)$ factor of $SO(4)=SU(2)_L\times SU(2)_R$.  
The subgroup of $SO(4)$ that commutes with $-M$ is then not $U(1)\times U(1)$
(as it is for generic $M$) but $SU(2)\times U(1)$.
As the dimension of the commuting subgroup of $-M$ increases by two when
(\ref{tono}) holds, the dimension of its orbit in $SO(4)$ drops by two.  So
while (\ref{tono}) puts a single real condition on the $v_i$, the space of
$M$'s for which it holds is of real codimension three.
In terms of $B$, this statement is easy to explain: for $B$ to be anti-self-dual
is three real conditions.

For the $D0\,$-$D6$ system, the situation is different.  A generic $-M$ commutes
with $U(1)\times U(1)\times U(1)\subset SO(6)$.  When (\ref{ono}) is obeyed,
the eigenvalues of $-M$ remain generically distinct and 
the stabilizer of $-M$ is still generically $U(1)\times U(1)\times U(1)$.
The dimension of the orbit does not jump, and so (\ref{ono}) places one real
condition on $M$ just as it does on the $v_i$.  The $D0\,$-$D6$ system
is supersymmetric on a locus of real codimension one in the space of
$B$-fields.  This is a crucial fact, for it will make it possible
later for the number of BPS states to jump in crossing this locus.

\bigskip\noindent{\it Analog For $D0\,$-$D2$ and $D0\,$-$D8$}

For the $D0\,$-$D2$ system, there is only a single rotation parameter
$v$.  The eigenvalues of $-M$ are $-e^{\pm 2\pi i v}$,  and the
determinant of $-M$ as an element of $U(1)$ is $-e^{\pm 2\pi i v}$
(with the sign depending on which $U(1)$ we pick). As $|v|<1/2$,
this cannot equal 1, so
turning on a $B$-field does not restore supersymmetry
for the $D0\,$-$D2$ system.  However, in the limit of $|B|\to\infty$,
one has $|v|\to 1/2$, and supersymmetry is restored.

For the $D0\,$-$D8$ system, there are four rotation parameters $v_i$.
The eigenvalues of $-M$ are $-e^{\pm 2 \pi i v_i}$.  The condition
for $-M$ to have determinant 1 in $U(4)$ is that, with some choice
of the signs,
\be
\pm v_1\pm v_2\pm v_3\pm v_4={\rm integer}.
\label{imico}
\ee
As $|v_i|<1/2$, there are two essentially different cases.
There is a supersymmetric locus with
\be
\pm v_1\pm v_2\pm v_3\pm v_4 = 0.
\label{timico}
\ee
This includes as a special case the possibility that $B_i=v_i=0$
for all $i$, which is the familiar supersymmetric $D0\,$-$D8$ system
without a $B$-field.  The more novel possibility is
\be
\pm v_1 \pm v_2 \pm v_3 \pm v_4 = \pm 1.
\label{himico}
\ee
Both the conventional supersymmetric locus and the novel one are
of real codimension 1 in the space of $B$-fields, since neither (\ref{timico})
nor (\ref{himico}) forces any degeneracies in the eigenvalues of $-M$.
Once again, we can make all of the signs positive with a suitable
choice of orientations.  When (\ref{timico}) or (\ref{himico}) is satisfied and the
$v_i$ are otherwise generic, there are two unbroken supersymmetries.


\section{BPS States}
\subsection{Analysis Of $D0\,$-$D6$ Spectrum}

We will  now  analyze the effective low energy physics of the supersymmetric
$D0\,$-$D6$ system.  We will show that this system supports a massless chiral
supermultiplet.  We will also show
 that the ground state energy of the $D0\,$-$D6$ system in the Neveu-Schwarz
sector changes sign as one crosses the supersymmetric locus.
(This was seen in \cite{chen}.)  We will then
argue, based on this, that on one side
of the supersymmetric locus  there is a supersymmetric or
BPS bound state, and on the other side there is not.

In the Ramond sector, we can analyze the situation as follows.
We work in light cone gauge with transverse oscillators $X^1,\dots,X^8$
and world-sheet superpartners $\psi^1,\dots,\psi^8$.   The ground
state energy vanishes in the Ramond sector  because of worldsheet supersymmetry.
In the notation of the last section, the worldsheet fermions $\psi^7$ and
$\psi^8$ (superpartners of $X^7$ and $X^8$, which describe motion orthogonal
to the $D6$-brane) have zero modes.  The other worldsheet fermions obey
twisted boundary conditions that depend on the $B$-field; for generic
$v_i$, they have no zero modes.  Quantization of the $\psi^7$ and $\psi^8$
zero modes give a pair of massless states whose spin (under rotations
of the 7-8 plane) is $\pm 1/2$.  Precisely one
 of these two states survives the GSO
projection.

Hence, for a supersymmetric configuration, there must also be a  massless
state in the Neveu-Schwarz sector, making part of a massless vector or chiral
supermultiplet.  Let us see how this comes about.
First we consider the case that all $v_i$ are positive with
\be
v_1+v_2+v_3=1/2.
\label{huvo}
\ee
We recall that a complex boson $X$ with modes $X_{n+1/2+\theta}$,
$|\theta|\leq 1/2$, has a ground state energy
\be
{1\over 24}-{\theta^2\over 2}.
\label{ncon}
\ee
A complex fermion $\psi$ with modes $\psi_{n+1/2+\theta}$ likewise
has a ground state energy
\be
-{1\over 24}+{\theta^2\over 2}.
\label{psicon}
\ee
In the problem at hand, for $X^1+iX^2$, $X^3+iX^4,$ and $X^5+iX^6$,
we have $\theta$ equal to $v_1,$ $v_2,$ and $v_3$, respectively.
For $X^7+iX^8$, $\theta=1/2$.  So the bosonic ground state energy
is
\be
\sum_{i=1}^3\left({1\over 24}-{v_i^2\over 2}\right)-{1\over 12}.
\label{bosgr}
\ee
For the fermions in the Neveu-Schwarz sector, the corresponding values
of $\theta$ are $v_i-1/2$ (for $i=1,2,3$) and 0.  So the fermionic ground
state energy is
\be
\sum_{i=1}^3\left(-{1\over 24}+{(1/2-v_i)^2\over 2}\right)
-{1\over 24}.
\label{fermgr}
\ee
Adding these up, we find the ground state energy to be
\be
{1\over 4}-{1\over 2}\sum_{i=1}^3v_i.
\label{inicni}
\ee

So in the supersymmetric case, with $\sum_iv_i=1/2$, the ground state
energy vanishes for $D0\,$-$D6$ strings in the NS sector.
There are no worldsheet
zero modes to be quantized, so there is a unique massless
$D0\,$-$D6$ state in this sector.  It has vanishing spin since, for example,
the filled fermi sea in this problem is invariant under a reflection
in the 7-8 plane.  Since we found a massless fermi state in the Ramond
sector, this unique massless state in the NS sector must survive the GSO
projection.  When $\sum_iv_i=1/2$, 
the massless bosonic and fermionic states from the NS and
R sectors combine with
similar states of $D6\,$-$D0$ strings to make a massless chiral multiplet.

Before analyzing the implications of this, let us first consider a more general
situation with unbroken supersymmetry.  There is no loss of essential
generality in picking positive signs in the supersymmetry condition
as in (\ref{ptono}), but we should not necessarily assume
that all $v_i$ are positive.  Since $|v_i|<1/2$, at least two $v_i$
must be positive if $v_1+v_2+v_3=1/2$.  So the remaining
essential case is that $1/2>v_1,v_2\geq 0$ and $0>v_3>-1/2$.  
These conditions imply
\be
|v_3|<v_1,v_2.
\label{polo}
\ee
With $v_3$ negative,
one must replace $v_3$ by $1-v_3$ in (\ref{fermgr}), and the ground state
energy comes out to be 
\be
-|v_3|.
\label{utolo}
\ee
So the NS sector is tachyonic in this case.  Spacetime supersymmetry
ensures that this tachyon is removed by the GSO projection.
The worldsheet fermions have modes $\psi_{n\pm v_i}$ (for $\psi^1,\dots,
\psi^6$) and $\psi_{n+1/2}$ (for $\psi^7,\psi^8$).  In view of (\ref{polo}),
this means that the lowest energy fermion creation operator has energy
precisely $|v_3|$.    Acting with this operator on the tachyonic ground
state, we get a unique GSO-even state of zero energy and zero spin.  This completes the demonstration
that in the supersymmetric case, quantization of the $D0\,$-$D6$
NS sector always gives a single massless scalar  state.

\subsection{ Bound State For $\sum_iv_i>1/2$}

Let us describe the low energy effective field theory of the $D0\,$-$D6$
system.  We think of it as a field theory -- really a quantum mechanical
system -- on the $D0$ world-line.

This theory has four unbroken supercharges -- the supercharges left
unbroken by the combined $D0\,$-$D6$ system for $\sum_iv_i=1/2$.
It has a $U(1)$ vector multiplet, which arises in quantizing the
0-0 strings.  Apart from the gauge field, this vector multiplet
contains fermions $\lambda$ and scalars $\vec \sigma$.
(One can think of the scalars as arising by dimensional reduction of
a four-dimenional vector multiplet to $0+1$ dimensions.
For a discussion of the possible low energy couplings of a vector multiplet
in $0+1$ dimensions, see \cite{diem}.)
The low energy theory 
also has a chiral multiplet $\Phi=\phi +\theta\psi+\dots$
that arises, as we have just seen, by quantizing the 0-6 and 6-0 strings.
Finally, there are three more massless chiral multiplets $T_\alpha$, $\alpha=1,\dots,3$
that arise from quantizing the 0-0 strings and describe the motion of the
$D0$-brane in the directions $x^1,\dots,x^6$ tangent to the $D6$-brane.
These  are Goldstone modes of an exact symmetry of spacetime
translations, and hence decouple from the low energy dynamics.

The physics of this system clearly depends very much on the
real parameter $r=\sum_iv_i-1/2$.  At $r=0$, supersymmetry is unbroken.
For $r<0$, the $D0\,$-$D6$ system is stable (as the NS ground state energy
is positive) but not supersymmetric.
The supersymmetry breaking is small if $|r|$ is very small, and it must
be possible to interpret it as spontaneous breaking of supersymmetry
in the low energy effective theory.  For $r>0$, the $D0\,$-$D6$ system 
is tachyonic and unstable.  It must decay to something else with the
same quantum numbers.  For small $r$, it is reasonable to expect to
be able to describe the stable state in the low energy effective theory.

With four supercharges, most coupling parameters are complex, and it is
somewhat unusual to see a real parameter playing an important role.
However, in this case, it is easy to see, by analogy with
known behavior of the $D0\,$-$D4$ system \cite{seibetal}, what effective field theory has the right properties:
the parameter $r$ is the Fayet-Iliopoulos coupling of the low energy
$U(1)$ gauge theory.  The potential energy function $V$ of the low energy
field theory then has a term proportional to $D^2$, where
$D=\bar\phi\phi -r$.  Thus
\be
V\sim (\bar\phi\phi-r)^2.
\label{unb}
\ee
Here we see that for $r<0$, the vacuum at $\phi=0$ is stable but
has positive energy and so describes a system with spontaneously
broken supersymmetry.  In this state, the $U(1)$ gauge symmetry is unbroken;
we interpret this state as the $D0\,$-$D6$ ground state.  For $r=0$, the $U(1)$
is still unbroken, and supersymmetry is also unbroken.  But for $r>0$,
there is a stable, supersymmetric vacuum with $|\phi|=\sqrt r$;
this vacuum has spontaneously broken $U(1)$ gauge symmetry, and the
phase of $\phi$ can be eliminated by a gauge transformation.  This
is the stable vacuum that arises from the $D0\,$-$D6$ system by tachyon
condensation when $r$ is slightly greater than zero.

Apart from the Goldstone multiplets $T_\alpha$ (which decouple from the low energy
dynamics), the state with $r>0$ and $|\phi|=\sqrt r$
is massive: all components of the vector and chiral multiplets have
mass in expanding around this vacuum. 
Hence, though we found this state by examination of
the classical potential, its existence 
is stable against quantum corrections.  There is no candidate for what
a Goldstone fermion might be in the event of supersymmetry breaking due
to quantum corrections.  (For $r<0$, the Goldstone fermion is part of the
vector multiplet, but for $r>0$, the vector multiplet gets mass by a Higgs
effect.)

The following question is puzzling at first sight.  For $r>0$, the
$D0\,$-$D6$ system, with the translations factored out, has a unique, massive,
supersymmetric vacuum, so that the supersymmetric index is $\Tr\,(-1)^F=1$.
For $r<0$, supersymmetry is spontaneously broken, so $\Tr\,(-1)^F=0$.  
How can $\Tr\,(-1)^F $ jump?  This jumping depends on the following
facts.  For $r\leq 0$, the effective $U(1)$ gauge theory also has a Coulomb
branch, in which the scalar fields $\vec \sigma$
in the vector multiplet get a vacuum expectation
value.  This Coulomb branch, of course, describes the motion of the 
$D0$ brane away from the $D6$ brane.  It is absent for $r>0$,
because there is a term $\vec\sigma^2|\phi|^2$ in the classical
potential; this term prevents $\vec\sigma$ from getting a vacuum expectation
value when $\phi$ has one.  (More informally, the supersymmetric ground
state for $r>0$ is a $D0\,$-$D6$ bound state, so it does not admit
any deformation to a Coulomb branch in which the two branes separate.)
For $r\leq 0$, because of the
$\vec\sigma^2|\phi|^2$ interaction, the field $\vec \sigma$
can only go to infinity at finite cost of energy if $\phi$ vanishes.
At this point, the classical energy is proportional to $r^2$,
because of (\ref{unb}).  If $\vec \sigma$ can go to infinity, the theory
has a continuous spectrum.
Hence, the theory has a continuous spectrum above a threshold proportional
to $r^2$.    At $r=0$, this threshold begins
at zero energy.  This means that at $r=0$, the counting of zero energy
states breaks down, and $\Tr\,(-1)^F$ can jump.

Though we have formulated our discussion for a $D0\,$-$D6$ system in
$\R^{10}$, the discussion is not much modified by compactification as
long as a sufficiently strong $B$-field can be turned on.  For
instance, we could consider compactification to $\R^4\times \T^6$ with
the $D6$-brane wrapped on $\T^6$.  Our discussion carries over to this
case directly; the supersymmetric
$D0\,$-$D6$ bound state that arises for $r>0$ is a finite
energy BPS state.
It is a $1/8$ BPS state -- invariant that is under 1/8 of the
32 supercharges of the toroidally compactified Type IIA superstring theory.  This bound state is absent for $r<0$,
showing explicitly that in string compactification to four dimensions
with 32 unbroken supersymmetries, the number of 1/8 BPS states
can jump as the parameters of the vacuum are modified.  Such jumping, which
might be important in some aspects of the counting of microscopic
states of BPS black holes in four dimensions, is analogous
to the jumping of 1/4 BPS states in ${\cal N}=4$ super Yang-Mills theory
in four dimensions \cite{bergman}.   

\subsection{ Generalization For Many Branes}

We can readily generalize these results to the case of $N$ $D0$-branes
interacting with $M$ coincident $D6$-branes.

The gauge group of the $D0$-branes is now $U(N)$.  The $D6$-branes
support a $U(M)$ gauge symmetry, which appears as a global symmetry
in the $D0$-brane quantum mechanics.  The quantization of the 0-6 strings
is the same as before, except that we must include Chan-Paton factors.
The upshot is that, on the supersymmetric locus, the massless 0-6 and 6-0
strings are a multiplet of massless chiral superfields $\Phi^i{}_k$, $i=1,\dots,N$,
$k=1,\dots,M$ transforming as $({ N},{ {\bar M}})$ of
$U(N)\times U(M)$.  

Allowing for the $U(N)$ Chan-Paton factors, the
massless 0-0 strings are, apart from the $U(N)$
vector multiplet, a trio of superfields $T_\alpha$ transforming in the
adjoint representation of $U(N)$.   In contrast to the $U(1)$ case,
where the $T_\alpha$ are neutral and decouple, for $U(N)$ they do not
decouple.   The $T_\alpha$ have a superpotential $W=Tr \,\,T_1[T_2,T_3]$,
which follows from dimensional reduction from ten-dimensional 
super-Yang-Mills theory to $0+1$ dimensions.

If we write $\Phi^i{}_k=\phi^i{}_k+\dots$ where the $\phi^i{}_k$ are
the $\theta=0$ components of the superfields, and likewise $T_\alpha
=t_\alpha+\dots,$ then the $D$-fields
are
\be 
D^i{}_j=\sum_k\phi^i{}_k\bar\phi^k{}_j+\sum_\alpha
[t_\alpha,\bar t_\alpha]^i{}_j-r\delta^i{}_j.
\label{kino}
\ee
(Here $\bar\phi^k{}_j$, $k=1,\dots,M$, $j=1,\dots,N$ transforms of course as 
$( {\bar N},{ M})$.)   A classical supersymmetric state is a solution
of $D^i{}_j=0$ together with $[t_\alpha,t_\beta]=0$ (the last condition
corresponds to $\partial W/\partial T_\alpha=0)$.  Since the trace of the $\phi\bar\phi$
term in $D $ is positive definite, while the $[t,\bar t]$ term
is traceless, such states exist
if and only if $r\geq 0$.  Moreover, if $r=0$, by taking the trace of
the equation $D^i{}_j=0$ we learn that $\phi^i{}_k=0$, so that the 0-6 strings
play no role and the discussion collapses to an analysis of the 0-0 strings
and their zero modes $t_\alpha$.\footnote{If the $\phi^i{}_k$ vanish
and $[t_\alpha,t_\beta]=0$, then the $D0$-branes are free to separate
from the $D6$-branes, by giving an expectation value to scalars in the
$U(N)$ vector multiplet.  Such configurations do not describe
$D0\,$-$D6$ bound states.}  So the
interesting case is $r>0$.  The moduli space ${\cal M}$ of classical 
supersymmetric ground states is the space of supersymmetric
classical states, divided by $U(N)$.  (Quantum supersymmetric ground
states must be found by studying a suitable quantum mechanics on ${\cal M}$.
We will not investigate this here.)

This description of the space of classical $D0\,$-$D6$ bound states is 
somewhat analogous
to the ADHM description of the moduli space of classical $D0\,$-$D4$ bound
states, that is, instantons, and the generalization of the ADHM construction to instantons on noncommutative ${\bf R}^4$, that is instantons with
a Fayet-Iliopoulos term
 \cite{nksh}.
One difference is that $D0\,$-$D4$ bound states are interesting both
in the absence of a Fayet-Iliopoulos term (classical instantons)
and in the presence of one (instantons on noncommutative space), while
for $D0\,$-$D6$, bound states exist only in the presence of a Fayet-Iliopoulos
term.  Moreover, of course, for $D0\,$-$D4$, there are bound states for
all values of $B$ (or all values for which $B$ is not anti-self-dual
if one has only one $D4$-brane), while for $D0\,$-$D6$, the $B$-field
must obey a certain inequality. Finally, by taking the trace of the equation
$D^i{}_j=0$, we learn that even for $r>0$, the $\phi^i{}_k$ are bounded
in absolute value, which corresponds roughly to an upper bound on the ``size''
of a bound state; there is of course no such bound for instantons.

It is perhaps of some interest for $r>0$ to consider the case that the $T_\alpha$
are all zero, corresponding roughly to all $D0$-branes being at the
same point in space.  Such solutions exist if and only if $M\geq N$.
For this case, the equation $D=0$ says that the $M$-component row vectors
whose components are $\phi^i{}_k/\sqrt r$ for fixed $i$ (the $M$ components
of the vector being labeled by $k=1,\dots,M$)
are orthonormal.  The space of such $N$-plets of orthonormal vectors,
modulo the action of $U(N)$, is the Grassmannian
$U(M)/U(N)\times U(M-N)$ of complex $N$-planes in ${\bf C}^M$.
So this is the moduli space of supersymmetric states with $T_\alpha=0$.

\subsection{Analog For $D0\,$-$D8$}

Now let us consider the analogous issues for the $D0\,$-$D8$ system.
We recall that there are two distinct supersymmetric loci of the $D0\,$-$D8$
system, governed respectively by (\ref{timico}) and (\ref{himico}).  
The first one contains as a special case the standard supersymmetric
$D0\,$-$D8$ system with $B=0$.  In this case, the ground state energy
in the NS sector is strictly positive, and there are no massless
bosons.  We therefore concentrate on the second supersymmetric locus.
Without essential loss of generality, we can pick orientations so
that the signs in (\ref{himico}) are all positive.  We moreover will
consider only the case that the $v_i$ are all positive, and leave the
interested reader to examine more general cases, by analogy with the
last paragraph of section 3.1.  Thus we assume
\be
0\leq v_i<1/2,~~\sum_{i=1}^4 v_i=1.
\label{pokko}
\ee
The analog of (\ref{bosgr}) for the bosonic contribution to the NS ground
state energy of the $D0\,$-$D8$ system is
\be 
\sum_{i=1}^4\left({1\over 24}-{v_i^2\over 2}\right),
\label{tuyo}
\ee
while the analog of (\ref{fermgr}) for the fermionic contribution is
\be
\sum_{i=1}^4\left(-{1\over 24}+{(1/2-v_i)^2\over 2}\right).
\label{buyo}
\ee
Adding these, the total ground state energy is
\be
{1\over 2}\left(1-\sum_iv_i\right).
\label{plopo}
\ee
Thus, we see the expected vanishing of the ground state energy when
$\sum_iv_i=1$.  
Moreover, for $\sum_iv_i<1$, the $D0\,$-$D8$ system is stable, though
not supersymmetric, and there is not evidence of a BPS bound state.
For $\sum_iv_i>1$, the $D0\,$-$D8$ system is unstable and presumably decays
to a supersymmetric or BPS ground state.  The effective low energy theory
is less constrained by supersymmetry, as there are only two supercharges.


\section{Comparison To Noncommutative Yang-Mills Theory}


The conditions for unbroken supersymmetry found in section 2 do not require
the $B$-field to be large in string units; it suffices for it to be of order 1.
However, while preserving unbroken supersymmetry, it is possible
to take $B\to\infty$ and thus to make contact with noncommutative
Yang-Mills theory.

In the limit that $b_i\to \pm \infty$, the rotation angles $v_i$ defined
in (\ref{hoko}) behave as
\be
v_i\to {1\over 2}{\rm sign}\,b_i-{1\over \pi b_i}.
\label{remigo}
\ee
In this limit, the supersymmetry condition (\ref{ono}) of 
the $D0\,$-$D6$ system becomes simply
\be
\pm {1\over b_1}\pm {1\over b_2}\pm {1\over b_3}=0.
\label{temigo}
\ee
Likewise, the supersymmetry condition (\ref{timico}) of the $D0\,$-$D8$ system
reduces to 
\be
\pm{1\over b_1}\pm{1\over b_2}\pm {1\over b_3}\pm{1\over b_4}=0.
\label{emigo}
\ee
We want to interpret these conditions in terms of noncommutative Yang-Mills
theory.

To do so, we should consider a suitable solution of noncommutative Yang-Mills
theory describing the $D0\,$-$D6$  system, find the
condition for this solution to be supersymmetric, and compare this
condition to (\ref{remigo}). (The $D0\,$-$D8$ system, and its comparison
to (\ref{emigo}), can be treated in precisely the same way.)
Which solution we want is quite
clear from several recent papers \cite{gn,polyc,mand,bak,stromet,harveyet,gntwo}.
To explain the situation starting from matrix theory (as in \cite{li,aoki,ishi,nishi,barm,ambj,alvw,fato,seiberg}),
we describe the $D6$-brane with a $B$-field via matrices $X^i$, $i=1,\dots,6$
obeying $[X^i,X^j]=i\theta^{ij}$.  This description is familiar
in matrix theory \cite{aa,bb,cc}. We likewise describe the $D0$-brane by
$1\times 1$ matrices with $X^i$ being multiples of the identity.  The $D0\,$-$D6$ brane solution
is described in matrix theory by taking the direct sum of the two
sets of matrices.  It can be reinterpreted as a solution of rank one
noncommutative
Yang-Mills theory  with 
\be
\hat F_{ij}= -\Pi\theta^{-1}_{ij},
\label{julp}
\ee
with $\Pi$ the projector onto a single state in Hilbert space
(this state originates with the $1\times 1$ matrices used to describe
the $D0$-brane).  For details, the reader may consult \cite{stromet,harveyet,gntwo}.

Next let us find the condition for this solution to be supersymmetric.
In general, the transformation law for the gluino field $\lambda$
in noncommutative super-Yang-Mills theory is $\delta\lambda
=\hat\Gamma^{ij}\hat F_{ij}\epsilon+\epsilon'$, where $\epsilon$ and $\epsilon'$
are respectively the unbroken and spontaneously broken supersymmetries.
Here $\hat\Gamma^{ij}={1\over 2}[\hat\Gamma^i,\hat\Gamma^j]$, where
\be
\{\hat\Gamma^i,\hat\Gamma^j\}=2G^{ij},
\label{furgo}
\ee
with $G^{ij}$ being the ``open string metric'' used in the noncommutative
super Yang-Mills theory.  The condition for unbroken supersymmetry
is 
\be
0=\hat F_{ij}\hat \Gamma^{ij}\epsilon +\epsilon'.
\label{tinon}
\ee
For the solution we are considering, $\hat F$ vanishes at infinity,
and hence we can
assume that $\epsilon'=0$.  Given (\ref{julp}), the condition for unbroken
supersymmetry thus reduces to
\be
\hat\Gamma^{ij}\theta^{-1}_{ij}\epsilon=0.
\label{gurgo}
\ee

To compare with the result of section 2, we must express this condition
in terms of the closed string variables -- the metric $g_{ij}$ and
$B$-field $B_{ij}$. In the following discussion, we will work entirely
in the six-dimensional spatial volume of the $D6$-brane (at fixed time)
with coordinates $x^1,\dots,x^6$.  We assume that $B$ is invertible in this
six-dimensional space; in fact, we assume that the ``eigenvalues''
$b_1,b_2,$ and $b_3$
are all tending to $\pm \infty$.
 In the $\alpha'\to 0$ limit,
the noncommutativity parameter $\theta$ and open string metric $G$ are \cite{sw}
\be
\theta=B^{-1},\,\, G_{ij}=-(2\pi \alpha')^2(Bg^{-1}B)_{ij}.
\label{uxxu}
\ee
We also introduce gamma matrices $\Gamma^i$ appropriate to the
closed string metric:
\be
\{\Gamma^i,\Gamma^j\}=2g^{ij},\,\Gamma_i=g_{ik}\Gamma^k,\,\,
\Gamma_{ij}={1\over 2}[\Gamma_i,\Gamma_j].
\label{puxxu}
\ee
Given the relation between $G$ and $g$, we can take the $\hat \Gamma^i$
to be 
\be
\hat \Gamma^i= (2\pi \alpha')^{-1}\theta^{ij}g_{jk}\Gamma^k.
\label{nuxxu}
\ee
The condition (\ref{gurgo}) of unbroken supersymmetry can hence be
expressed in terms of closed string variables:
\be
\Gamma_{ij}(B^{-1})^{ij}\epsilon=0.
\label{upuxxu}
\ee

If $B$ is the direct sum of $2\times 2$ blocks with ``eigenvalues'' $b_1,b_2,b_3$, then
$B^{-1}$ is likewise a direct sum of $2\times 2$ blocks with
eigenvalues $-b_i^{-1}$, $i=1,2,3$.  So (\ref{upuxxu})  becomes
\be
\left(b_1^{-1}\Gamma_{12}+b_2^{-1}\Gamma_{34}
+b_3^{-1}\Gamma_{56}\right)\epsilon=0.
\label{nupuxxu}
\ee
The matrices $\Gamma_{12}$, $\Gamma_{34}$, and $\Gamma_{56}$ commute
and have eigenvalues $\pm i$, with each set of signs $(\pm i,\pm i,\pm i)$
arising for some eigenvector.  So the condition that (\ref{nupuxxu}) is obeyed
for some $\epsilon$ is that
\be
\pm b_1^{-1}\pm b_2^{-1}\pm b_3^{-1}=0,
\label{tupuxxu}
\ee
with some set of the signs.  This is in full agreement with (\ref{temigo}).


\acknowledgments
This work was supported in part by NSF Grant PHY-9513835 and the Caltech Discoverty Fund.  I would like to thank J. Gomis for discussions.



\begin{thebibliography}{99}

\bibitem{polch} J. Polchinski, {\it Superstring Theory}, vol. 2, chapter 13
 (Cambridge University Press, 1998).

\bibitem{sw} N. Seiberg and E. Witten, ``String Theory And Noncommutative 
Geometry,'' hep-th/9908142, JHEP {\bf 9909} (1999) 032.
 
\bibitem{chen} B. Chen, H. Itoyama, T. Matsuo,
and K. Murakami, ``$p-p'$ System With $B$-field, Branes at Angles,
And Noncommutative Geometry,'' hep-th/9910263.

\bibitem{inyong} M. Mihalescu, I. Y. Park, and T. A. Tran,
``$D$-Branes As Solitons Of An $N=1$, $D=10$ Non-commutative Gauge Theory,''
hep-th/0011079.

\bibitem{bergman} O. Bergman, ``Three Pronged Strings And 1/4 BPS States in
${\cal N}=4$ Super-Yang-Mills Theory,'' hep-th/9712211;
O. Bergman and B. Kol, ``String Webs And 1/4 BPS Monopoles,''
hep-th/9804160, Nucl. Phys. {\bf B536} (1998) 149.

\bibitem{cds} A. Connes, M. R. Douglas, and A. Schwarz,
``Noncommutative Geometry And Matrix Theory: Compactification On Tori,''
JHEP {\bf 9802} (1998) 003.

\bibitem{gn} D. J. Gross and N. A. Nekrasov, ``Monopoles And Strings In
Noncommutative Gauge Theory,'' hep-th/0005204, JHEP {\bf 0007} (2000) 034,
``Dynamics Of Strings in Noncommutative Gauge Theory,'' hep-th/0007204,
JHEP {\bf 0010} (2000) 021.

\bibitem{polyc} A. P. Polychronakos, ``Flux Tube Solutions In Noncommutative
Gauge Theories,'' hep-th/0007043.

\bibitem{mand} D. P. Jatkar, G. Mandal, and S. R. Wadia, ``Nielsen-Olesen
Vortices In Noncommutative Abelian Higgs Model,'' hep-th/0007078,
JHEP {\bf 0009:018} (2000).

\bibitem{bak} D. Bak, ``Exact Solutions Of Multi-Vortices And False
Vacuum Bubbles In Noncommutative Abelian-Higgs Theories,''
hep-th/00008204.

\bibitem{stromet} M. Aganagic, R. Gopakumar, S. Minwalla, and A. Strominger,
``Unstable Solitons In Noncommutative Gauge Theory,'' hep-th/0009142. 

\bibitem{harveyet} J. A. Harvey, P. Kraus, and f. Larsen, ``Exact 
Noncommutative Solitons,'' hep-th/0010060.

\bibitem{gntwo} D. J. Gross and N. A. Nekrasov, ``Solitons In Noncommutative
Gauge Theory,'' hep-th/0010090.

\bibitem{kachru} S. Kachru and J. McGreevy, ``Supersymmetric Three-cycles And (Super)symmetry
Breaking,' hep-th/9908135, Phys. Rev. {\bf D61} (2000) 026001. 

\bibitem{douglas} M. Berkooz, M. R. Douglas, and R. G. Leigh, ``Branes 
Intersecting At Angles,'' hep-th/960139, Nucl. Phys. {\bf B480}
(1996) 265.

\bibitem{tf} E. S. Fradkin and A. A. Tseytlin, ``Nonlinear Electrodynamics
>From Quantized Strings,'' Phys.l Lett. {\bf 163B} (1985) 123.

\bibitem{napetal} C. G. Callan, C. Lovelace, C. R. Nappi, and S. A. Yost,
``String Loop Corrections To Beta Functions,'' Nucl. Phys. {\bf B288} (1987)
525; A. Abouelsaood, C. G. Callan, C. R. Nappi, and S. A. Yost,
``Open Strings In Background Fields,'' Nucl. Phys. {\bf B280} (1987) 599.

\bibitem{diem} E. Diaconescu and R. Entin,
``A Nonrenormalization Theorem For The $D=1$, ${\cal N}=8$ Vector
Multiplet,'' hep-th/9706059, Phys. Rev. {\bf D56} (1997) 8045.

\bibitem{seibetal} O. Aharony,
M. Berkooz, and N. Seiberg, ``Light Cone Description Of $(2,0)$
Superconformal Theories In Six-Dimensions,'' Adv. Theor. Math. Phys.
{\bf 2} (1998) 119.


\bibitem{nksh} N. Nekrasov and A. Schwarz, ``Instantons on Noncommutative
${\bf R}^4$ and $(2,0)$ Superconformal Six-Dimensional Theory,''
hep-th/9802068, Commun. Math. Phys. {\bf 198} (1998) 689.

\bibitem{li} M. Li, ``Strings From IIB Matrices,'' hep-th/9612222,
Nucl. Phys. {\bf B499} (1997) 149.

\bibitem{aoki} H. AOki, N. Ishibashi, S. Iso, H. Kawai, Y. Kitazawa,
and T. Tada, ``Noncommutative Yang-Mills In IIB Matrix Model,'' hep-th/9908141,
Nucl. Phys. {\bf B565} (2000) 176.

\bibitem{ishi} N. Ishibashi, ``A Relation Between Commutative And
Noncommutative Descriptions Of $D$-Branes,'' hep-th/9909176.

\bibitem{nishi} N. Ishibashi, S. Iso, H. Kawai, and Y. Kitazawa,
``Wilson Loops In Noncommutative Yang-Mills,'' hep-th/9910004.

\bibitem{barm} I. Bars and D. Minic, ``Noncommutative Geometry On A Discrete
Periodic Lattice And Gauge Theory,'' hep-th/9910091.

\bibitem{ambj} J. Ambjorn, Y. M. Makeenko, J. Nishimura, and R. Szabo,
``Finite $N$ Matrix Models Of Noncommutative Gauge Theory,''
hep-th/9911041, JHEP {\bf 11} (1999) 029, ``Nonperturbative Dynamics
Of Noncommutative Gauge Theory,'' hep-th/0002158; J. Ambjorn,
K. N. Anagnostopoulos, W. Bietenholz, T. Hotta, and J. Nishimura,
``Large $N$ Dynamics Of Reduced 4D $SU(N)$ Super Yang-Mills Theory,''
hep-th/0003208,
JHEP {\bf 0007} (2000) 013; J. Ambjorn, Y. M. Makeenko, J. Nishimura,
and R. Szabo, ``Lattice Gauge Fields And Discrete Noncommutative Yang-Mills
Theory,'' hep-th/0004147, JHEP {\bf 0005} (2000) 003.
 
\bibitem{alvw} L. Alvarez-Gaum\'e and S. R. Wadia, ``Gauge Theory On
A Quantum Phase Space,'' hep-th/0006219.

\bibitem{fato} A. H. Fatollahi, ``Gauge Symmetry As Symmetry Of Matrix
Coordinates,'' hep-th/0007023.

\bibitem{seiberg} N. Seiberg, ``A Note On Background Independence In
Noncommutative Gauge Theories, Matrix Model and
Tachyon Condensation,'' hep-th/0008013, JHEP {\bf 0009} (2000) 003.

\bibitem{aa} T. Banks, W. Fischler, S. H. Shenker, and L. Susskind,
``$M$ Theory As A Matrix Model: A Conjecture,'' hep-th/9610043,
Phys. Rev. {\bf D55} (1997) 5112.

\bibitem{bb} O. J. Ganor, S. Ramgoolam, and W. I. Taylor, ``Branes, Fluxes,
and Duality In M(atrix)-Theory,''hep-th/9611202, Nucl. PHys. {\bf B492} (1997)
191.

\bibitem{cc} T. Banks, N. Seiberg, and S. Shenker, ``Branes From Matrices,''
hep-th/9612157, Nucl. Phys. {\bf B490} (1997) 91.

\end{thebibliography}
\end{document}